\documentclass[%
 preprint,
superscriptaddress,
longbibliography,
 amsmath,amssymb,
 aps,
 pra,
]{revtex4-1}

\usepackage{graphicx}
\usepackage{appendix}
\usepackage{color}
\usepackage{dcolumn}
\usepackage{bm}
\usepackage{ulem}

\begin{document}
\title{Microstructural and rheological training and memory of nanocolloidal soft glasses under cyclic shear}

\author{Yihao Chen}
\affiliation{%
Department of Physics and Astronomy, Johns Hopkins University, Baltimore, Maryland 21218, USA
}%
\author{Simon A. Rogers}%
\affiliation{%
Department of Chemical and Biomolecular Engineering, University of Illinois Urbana-Champaign, Champaign, IL 61801, USA
}%
\author{Suresh Narayanan}
\affiliation{X-Ray Science Division, Argonne National Laboratory, Argonne, Illinois 60439, USA}
\author{James L. Harden}
\affiliation{%
 Department of Physics, University of Ottawa, Ottawa, Ontario K1N 6N5, Canada
}%
\author{Robert L. Leheny}
\affiliation{%
Department of Physics and Astronomy, Johns Hopkins University, Baltimore, Maryland 21218, USA
}%

\date{\today}

\begin{abstract}
  An intrinsic feature of disordered and out-of-equilibrium materials, such as glasses, is the dependence of their properties on their history.  An important example is rheological memory, in which disordered solids obtain properties based on their mechanical history.  Here, we employ x-ray photon correlation spectroscopy (XPCS) with \textit{in situ} rheometry to characterize memory formation in a nanocolloidal soft glass due to cyclic shear.  During a cycle, particles undergo irreversible displacements composed of a combination of shear-induced diffusion and strain fields.  The  magnitudes of these displacements decrease with each cycle before reaching a steady state where the microstructure has become trained to achieve enhanced reversibility. The displacements resemble a random walk in which the directions in each cycle are independent of those in preceding cycles.  Accompanying the training is a steady decrease in the dissipation during each cycle towards a steady state value.  Memory of this training is revealed by measurements in which the amplitude of the shear is changed after steady state is reached. The magnitude of the particle displacements as well as the dissipation and the change in residual stress vary non-monotonically with the new strain amplitude, having minima near the training amplitude, thereby revealing both microscopic and macroscopic signatures of memory.
\end{abstract}

\maketitle

\section{Introduction}

Modifying a material's properties through mechanical work has been an important processing strategy for centuries.  Recently, appreciation has grown for how such dependence on mechanical history is an intrinsic attribute of materials that are out-of-equilibrium and further how investigations of the process provide a unique window into the aquisition and retention of memory in such systems~\cite{KeimRMP2019}.  The behavior of amorphous solids under cyclic shear, where microscopic particle rearrangements evolve with repeated shearing, has become a canonical example of the ability of out-of-equilibrium materials to encode mechanical memory~\cite{FioccoPRL2014,FioccoJPCM2015,KeimPRL2014,KeimSciAdv2021,FarhadiSM2017,AdhikariEPJE2018,MukherjiPRL2019,PriezjevPRE2017,BensonPRE2021,SchwenSofMat2020,PaulsenPRL2014} that can be revealed later by a reading protocol~\cite{FioccoPRL2014,FioccoJPCM2015,AdhikariEPJE2018,PaulsenPRL2014}. Such memory formation has been studied in detail in experiments on athermal and two-dimensional (2D) colloidal glasses formed at an interface~\cite{KeimPRL2014,MukherjiPRL2019,KeimPRR2020,PaulsenPRL2014} and in simulations of glasses~\cite{KeimPRE2013,FioccoPRL2014,FioccoJPCM2015,LundbergPRE2008,PriezjevPRE2017,ArceriPRE2021,LindemanSciAdv2021}. The key signature of the memory observed in these systems is the reduction in irreversible displacements that particles experience with increasing number of strain cycles.  Significantly, the trained glasses can possess so-called loop reversibility, where the particles follow different paths during each half of a cycle of strain, or even over multiple cycles, but ultimately have no net displacement ~\cite{LundbergPRE2008,SchreckPRE2013,KeimPRL2014,LavrentovichPRE2017,GallowaySM2020,KeimSciAdv2021,SzulcJCP2022,MunganPRL2019,TerziPRE2020}.  This behavior instills in the systems a memory of the specific strain amplitude at which they were trained.  

Despite this previous work, significant questions remain regarding the process of training glasses through cyclic shear and the memory that it encodes.  One question is whether and in what ways bulk (3D), thermal glasses exhibit such memory formation.  Also, while numerous studies have characterized memory in terms of microscopic reversibility, far less is known about how memory of cyclic shear might become encoded and read through measurements of a glass's macroscopic mechanical properties~\cite{ChattopadhyayJCP2022}.
Here, we address these questions in experiments employing x-ray photon correlation spectroscopy with {\it in situ} rheometry (rheo-XPCS) to investigate the evolution of a nanocolloidal soft glass's microscopic dynamics and macroscopic rheology simultaneously during the start-up of cyclic shear.  XPCS, which functions similarly to dynamic light scattering (DLS) but accesses nanoscale dynamics, uses time correlations in the scattering of a coherent beam to probe temporal changes in microstructure.  When combined with {\it in situ} mechanical testing, XPCS has been shown to be effective in characterizing stress-induced dynamics in a variety of contexts~\cite{LehenyCurOpn2015,BuschEPJE2008,UrbaniJSR2016,Westermeier2016,ChenPRM2020,ChenPOF2021,SuttonPRR2021,PrestoACSAMI2023,DonleyPNAS2023}. Both XPCS and DLS have been used to investigate microscopic irreversibility under cyclic shear in glasses and gels~\cite{HebraudPRL1997, HohlerPRL1997, PetekidisPRE2002, SmithPRE2007, LauratiJRheol2014, RogersPRE2014, RogersPRM2018, EderaSoftMatter2021}, but these previous studies focused on the behavior of well trained systems where the microstructure had already adapted to the shear deformation.  In this work, we focus on the evolving microstructural response of the glass during the onset of cyclic shear to gain insight into the training process and on the effects of this training as a form of memory encoded in both the microscopic properties and macroscopic rheology of the glass. 

\begin{figure*}
    \centering
    \includegraphics[width = 6.5in]{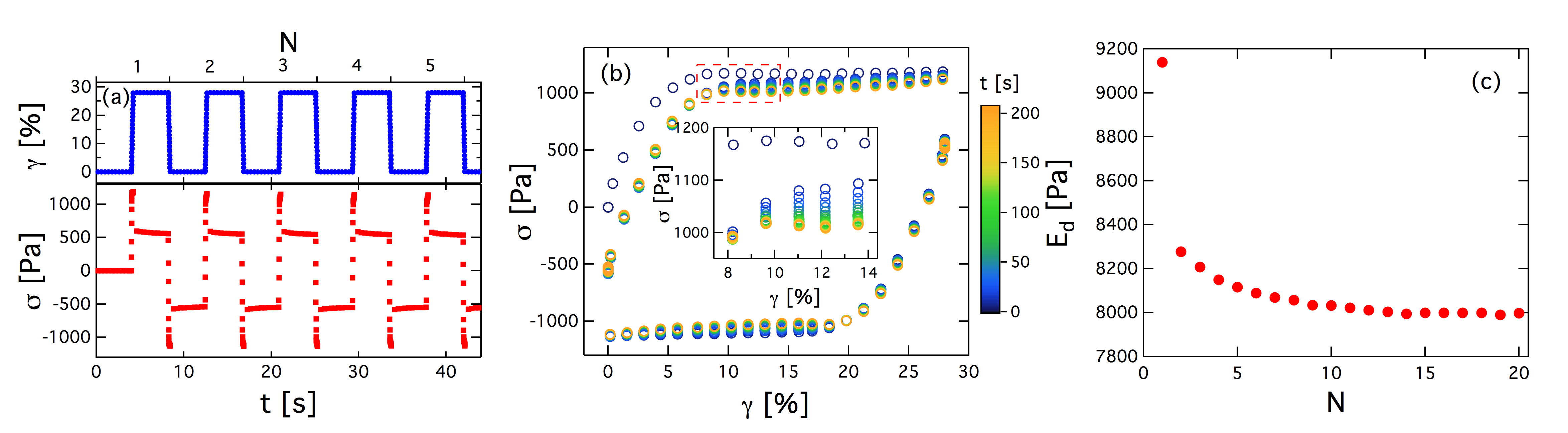}
    \caption{\textbf{Strain profile and rheological response to cyclic shear.} (a) Applied strain (top panel) and corresponding stress (bottom panel) as a function of time during the first 5 cycles of shear with $\gamma_{\text{train}}$ = 28\% and period $T=$ 8.4 s. (b) Stress as a function of strain during the first 24 cycles of the measurement. The colorbar indicates the time since the start of the measurement.  Inset: An enlarged version of the region indicated by the red square in main panel. (c) Dissipation in a cycle, as obtained from the area enclosed by stress-strain loop, as a function of cycle number.}
    \label{figTrainingrheo}
\end{figure*}

\section{Training}
\subsection{Rheological Training}
Experiments on the nanocolloidal soft glass employed a square-wave strain profile between $\gamma = 0$ and $\gamma_\text{train}$ with period $T = 8.4$ s, as illustrated in the upper panel of Fig.~\ref{figTrainingrheo}(a) for $\gamma_\text{train}$ = 28\%.  In each cycle, the strain increased from 0 to $\gamma_\text{train}$ at a constant rate in 0.2 s.  The strain was then fixed at $\gamma_\text{train}$ for 4 s before returning to 0, again at a constant rate in 0.2 s.  The strain then remained fixed at 0 for 4 s before the cycle repeated.  The glass's reological response and corresponding shear-induced microscopic dynamics due to the cyclic shear were characterized in rheo-XPCS measurements with $\gamma_\text{train}$ ranging from 1\% to 56\%.  An example of the glass's rheological behavior is shown in Fig.~\ref{figTrainingrheo}(b), which displays the stress-strain response during the first 24 cycles of strain at $\gamma_\text{train}$ = 28\%.  As the strain rises from zero, the stress initially increases, indicating a primarily elastic response.  Above $\gamma \approx $ 10\%, the stress becomes largely independent of strain, indicating yielding and viscoplastic flow.  Further details regarding the yielding behavior are provided in the Supplementary Material.  When the strain decreases from $\gamma_\text{train}$ to 0, the process repeats, with the glass initially showing primarily elastic response before yielding and undergoing viscoplastic flow.  During each interval of constant strain at $\gamma = 0$ or $\gamma = \gamma_\text{train}$, the stress decreases with time before reaching a plateau residual stress.  This time dependence is illustrated in the lower panel of Fig.~\ref{figTrainingrheo}(a), which shows the stress as a function of time during the first few cycles of strain at $\gamma_\text{train} =$ 28\%.  The stress is near zero in the first interval of zero strain. As the strain ramps to $\gamma_\text{train}$ = 28\%, the stress first increases and then reaches a plateau near 1200 Pa where the glass experiences yielding. When the strain is held at 28\%, the glass undergoes stress relaxation with a rapid drop to about 600 Pa followed by a slow decrease.  A detailed rheo-XPCS study of stress relaxation at fixed strain is presented in Ref.~\cite{ChenPRM2020}.  

After the strain ramps back to zero, the stress again experiences yielding and relaxation processes, but in the opposite direction. More importantly, the stress becomes negative and significantly deviates from zero after the first cycle.  Thus, although the strain profile is asymmetric about zero, the stress response becomes symmetric about zero.  Significantly, the stress response of the glass does not simply repeat with each subsequent strain cycle but steadily evolves during the first several cycles toward a steady state.  This evolution is illustrated by the inset of Fig.~\ref{figTrainingrheo}(b), which shows the stress in the region demarkated by the red box on an expanded scale.  As shown in Fig.~\ref{figTrainingrheo}(c), the dissipation $E_d$, obtained from the area within the stress-strain loops, decreases steadily during approximately the first 15 cycles of strain.  Thus, the cyclic shear modifies the glass microstructure such that the rheological response becomes less dissipative.  

\subsection{Microstructural Training}
Insights into the microstructural changes induced by the cyclic shear are obtained from the XPCS measurements through analysis of the 
 instantaneous correlation function~\cite{MadsenNJP2010}
\begin{equation}
    C(\textbf{q},t_1,t_2) = \frac{<I(\textbf{q},t_1)I(\textbf{q},t_2)>}{<I(\textbf{q},t_1)><I(\textbf{q},t_2)>}
    \label{InsCorrFun}
\end{equation}
where $I(\textbf{q},t)$ is the coherent scattering intensity at wave-vector $\textbf{q}$ and time $t$, and the brackets represent averages over detector pixels in a small vicinity centered around $\textbf{q}$.  Our analysis of the XPCS measurements focuses on $\textbf{q}$ along the vorticity direction of the shear to highlight the three-dimensional nature of the shear-induced particle motion.  An example of results for $\textbf{q}$ in the flow direction is shown in the Supplementary Materials. Figure~\ref{figTrainingCorr}(a) shows a colormap of $C(\textbf{q},t_1,t_2)$ at wave-vector $q$ = 0.37 nm$^{-1}$, which is near the interparticle structure factor peak (see Supplementary Materials) for $\gamma_\text{train}$ = 1\%. The colormap is symmetric about the diagonal ($t_1 = t_2$) and consists of alternating squares of high (yellow) and low (blue) correlation regions. The squares of high correlation along the diagonal correspond to time intervals when the macroscopic strain is held fixed. Due to dynamical arrest in the glass, the microstructure remains largely unchanged within each interval of fixed macroscopic strain. As the difference between $t_1$ and $t_2$ increases such that $t_1$ and $t_2$ fall into adjacent intervals with different macroscopic strain, the correlation dramatically drops due to the particle displacements induced by the strain~\cite{BurghardtPRE2012,LehenyCurOpn2015}, leading to the squares of low correlation centered at half a period from the diagonal ($|t_1 - t_2| = T/2$). As the difference between $t_1$ and $t_2$ increases further, the times fall into intervals at the same macroscopic strain separated by one cycle, and the particles have nearly no net displacement resulting in squares of high correlation centered at one period from the diagonal  ($|t_1 - t_2| = T$). Similarly, high correlation regions are centered at two periods from the diagonal ($|t_1 - t_2| = 2T$), and low correlation region are centered  at 3/2 and 5/2 periods from the diagonal.  $C(\textbf{q},t_1,t_2)$  at times separated by full strain cycles is nearly as large as within the same interval of fixed strain, indicating the deformation due to $\gamma_\text{train}$ = 1\% is essentially fully reversible, which is reasonable given that the strain amplitude is well within the glass's regime of linear elastic response.


\begin{figure*}
    \centering
    \includegraphics[width = 6.5in]{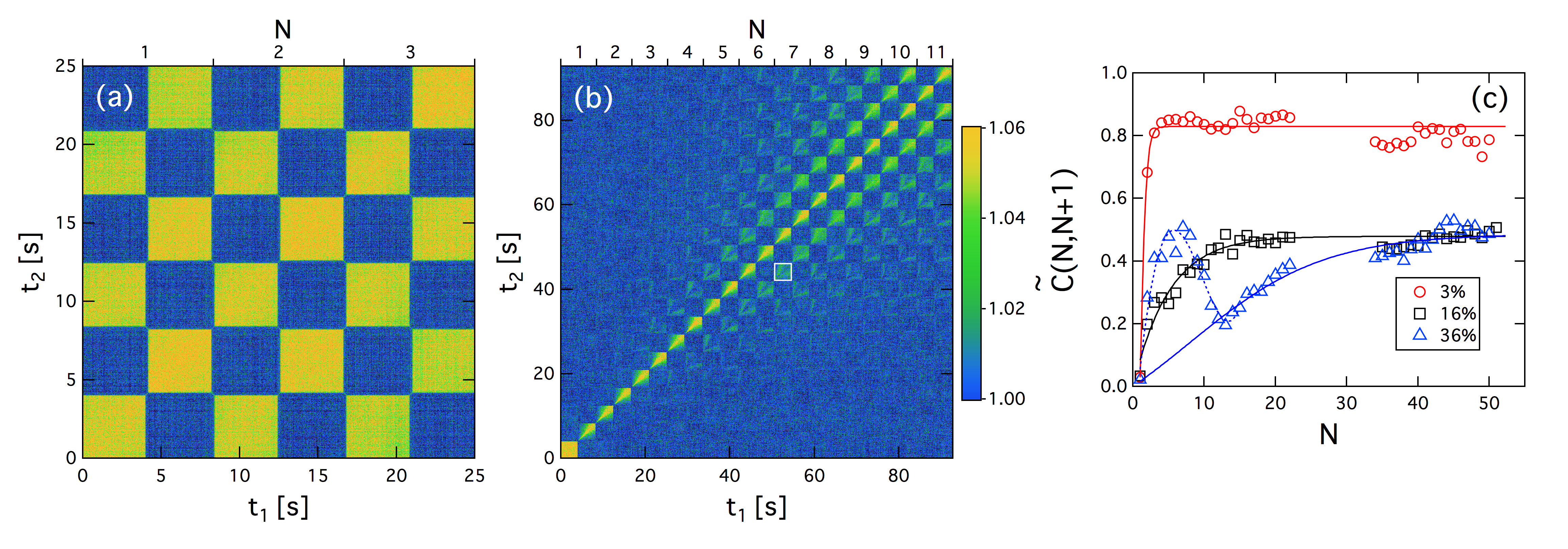}
    \caption{\textbf{Microstructural response to cyclic shear.} (a) Instantaneous correlation function $C(\textbf{q},t_1,t_2)$ during the first three cycles of a measurement with $\gamma_\text{train}$ = 1\%. (b) $C(\textbf{q},t_1,t_2)$ during the first 11 cycles of a measurement with $\gamma_\text{train}$ = 8\%. The white square indicates the region over which the average in the numerator of Eq.~\eqref{eq_normalizedC} is taken for $N$ = 6. (a) and (b) share the correlation colorbar. (c) Normalized correlation function $\tilde{C}(\textbf{q};N,N+1)$ as a function of cycles during measurements with different $\gamma_\text{train}$, as specified in the legend. Solid lines are the results of fits using Eq.~\eqref{eq_fitCorrVSN}. The dotted line is a guide to the eye.  In all cases, $q$ = 0.37 nm$^{-1}$ in the vorticity direction. }
    \label{figTrainingCorr}
\end{figure*}

Figure~\ref{figTrainingCorr}(b) shows the colormap of $C(\textbf{q},t_1,t_2)$ during the first 11 cycles with $\gamma_\text{train}$ = 8\%, which is beyond the linear elastic region and near the yield point. High correlation regions centered along the diagonal ($t_1 = t_2$) during times of fixed strain are again visible, although the correlations within the squares show fine structure reflecting dynamics of stress relaxation at the fixed strain~\cite{ChenPRM2020}.  However, the correlations in microstructure at different time intervals corresponding to the same macroscopic strain state show an evolution during the start-up of the cyclic shear. For example, the correlation of microstructures separated by one full cycle, represented by the squares centered one period from the diagonal ($|t_1 - t_2| = T$), is small for the first few cycles and gradually increases with the number of cycles. This trend is also seen in the microstructures separated by multiple  cycles ($|t_1 - t_2|$ = 2$T$, 3$T$, 4$T$, etc.). This evolution indicates increasing microscopic reversibility with each cycle of shear.

To quantify the correlations between microstructures separated by one cycle, we introduce a normalized correlation function, 
\begin{equation}
    \tilde{C}(\textbf{q};N,N+1) = \frac{<C(\textbf{q},t_1,t_2)>_{t_1,t_2}-1}{<C(\textbf{q},t',t'+\Delta t)>_{t'}-1}
    \label{eq_normalizedC}
\end{equation}
where the brackets in the numerator represent averages over $t_2$ and $t_1$ in the first 4 s of the $N$th and $(N+1)$th cycle, respectively.  For example, the white square in Fig.~\ref{figTrainingCorr}(b) indicates the region over which the numerator of is averaged for $N =$ 6.  The brackets in the denominator represent an average over $t'$ in the first 4 s of both the $N$th and $(N+1)$th cycles, and $\Delta t$ = 0.02 s is the time between adjacent x-ray images. This term is meant to account for contributions from thermal fluctuations.  Specifically, particles in the glass undergo rapid, thermally driven, ``caged'' motion that leads to a suppression of $C(\textbf{q},t_1,t_2)$ below its instrumental limit ({\it{i.e.}}, the Siegert factor) at the shortest accessible time difference $|t_1 - t_2|$ by an amount that depends on wave vector~\cite{BandyopadhyayPRL2004,SchweizerJCP2007}.  (For details, see the Supplementary Material.)  Normalized in this way to account for the thermal effects, $\tilde{C}$ hence quantifies the shear-induced differences in microstructure before and after the $N$th cycle.  

Figure~\ref{figTrainingCorr}(c) shows $\tilde{C}$ at $q$ = 0.37 nm$^{-1}$ as a function of cycle number for $\gamma_{\text{train}}$ = 3, 16, and 36\%, illustrating three types of behavior observed.  
At $\gamma_\text{train}$ = 3\%, which is just above the regime of linear response (see Fig.~S3 in the SI.), $\tilde{C}$ increases very quickly to a large plateau value that indicates almost full reversibility between cycles.  At $\gamma_\text{train}$ = 16\%, which is near yielding, $\tilde{C}$ starts close to zero and gradually rises to a plateau at a reduced value, implying an increasing similarity between microstructures and less rearrangement after each successive cycle.  At an even larger strain, $\gamma_\text{train}$ = 36\%, which is well above yielding, $\tilde{C}$ shows a non-monotonic evolution in which it increases then decreases temporarily before rising again toward an apparent plateau.  This non-monotonic evolution is remarkably similar to that seen in recent simulations of glasses under cyclic shear in which the average potential energy and average displacement of the particles were tracked as a function of cycle number~\cite{ParmarPRX2019}.  In the simulations, the non-monotonic behavior of these quantities was identified with the formation of a shear band and the resulting competing effects of increasing energy and displacements within the band and decreasing energy and displacements outside it.  We speculate that such spatially heterogeneous fluidization, like with shear band formation, similarly occurs in the experiements at sufficiently high strain amplitude, leading to the observed non-monotonic behavior.  This hypothesis is supported by the observation of training up to the largest applied strain, $\gamma_\text{train}$ = 56\%, where the ability to train the microstructure for enhanced reversibility is very unlikely in the absence of a heterogenous strain profile.

\begin{figure*}
    \centering
    \includegraphics[width = 6.5in]{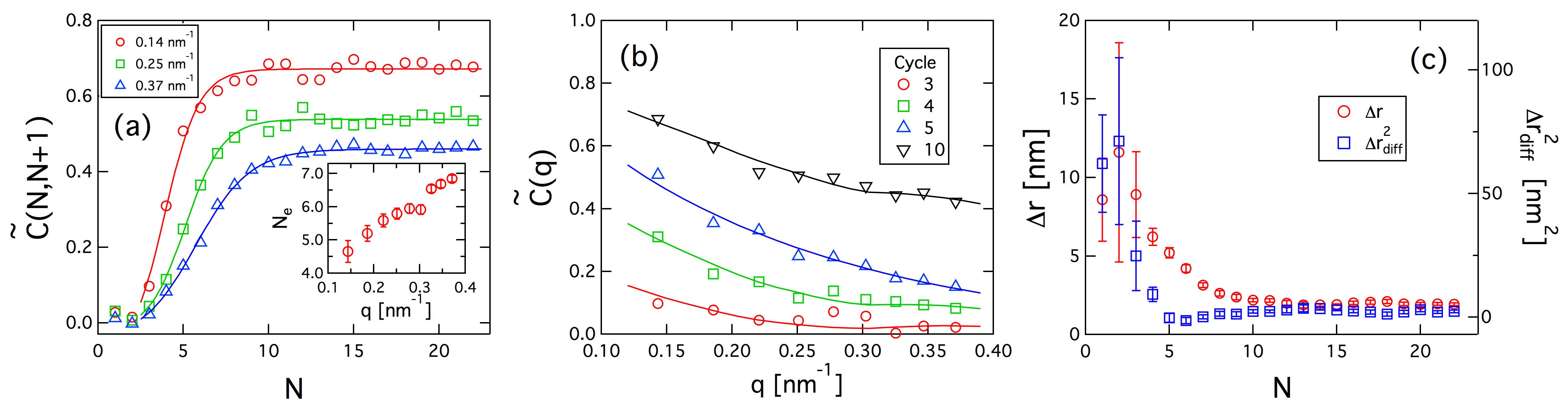}
    \caption{\textbf{Evolution in microstructural changes during cyclic shear.}  (a) $\tilde{C}(\textbf{q};N,N+1)$ at three wave vectors in the vorticity direction as a function of cycle number during a  measurement with $\gamma_\text{train}$ = 8\%. Solid lines are the results of fits using Eq.~\eqref{eq_fitCorrVSN}. Inset: Training time (measured in number of cycles) as a function of wave vector. (b) $\tilde{C}(q;N,N+1)$ as a function of wave vector following different numbers of cycles, as specified in the legend, during the measurement with $\gamma_\text{train}$ = 8\%. Solid lines show fits to the data using Eq.~\eqref{eq_fitmodel}. (c) $\Delta r$ and $\Delta r_\text{diff}^2$ as functions of the cycle number for $\gamma_\text{train}$ = 8\%.}
    \label{figCorrFit}
\end{figure*}

Figure~\ref{figCorrFit}(a) shows $\tilde{C}$ at three wave vectors as a function of cycle number for $\gamma_\text{train}$ = 8\%.  $\tilde{C}$ rises faster and reaches a higher plateau at smaller wave vector, incidating that the microstructure trains more quickly and achieves greater reversibility when viewed on larger length scales. The sigmoidal shape of $\tilde{C}$ in Figs.~\ref{figTrainingCorr}(c) and \ref{figCorrFit}(a) can be captured by the empirical form
\begin{equation}
    \tilde{C}(N,N+1) = A\tanh(cN + D) + B
    \label{eq_fitCorrVSN}
\end{equation}
where $A$, $B$, $c$, and $D$ are $q$-dependent fitting parameters, as shown by the lines in the figures. The training time $N_e$, defined as the number of cycles required for $\tilde{C}$ to reach $(1 - 1/e)$ of its plateau and shown at $\gamma_\text{train}$ = 8\% in the inset to Fig.~\ref{figCorrFit}(a), increases approximately linearly with wave vector.


The dependence of $\tilde{C}$ on wave vector results from the nature of the shear-induced irreversible displacements, which we model as a superposition of diffusive motion and heterogeneous strain fields. The correlation function due to these two kinds of rearrangements has a form
\begin{equation}
    \tilde{C}(q;N,N+1) = \exp\left[-\frac{q^2}{3S_M(q)}\Delta r_\text{diff}^2\right] \exp[ - q\Delta r]
    \label{eq_fitmodel}
\end{equation}
where $\Delta r_\text{diff}^2$ $\Delta r$ are the $N$-dependent the mean square shear-induced diffusive displacement and characteristic strain displacement, respectively, and $S_M(q)$ is the measurable structure factor of the soft glass obtained through small angle x-ray scattering (SAXS) measurements on the quiescent sample~\cite{Scheffold_2009}.  The diffusion term is modeled after the non-ergodicity parameter for glasses that characterizes localized thermal diffusion, where the inclusion of $S_M(q)$ accounts for effects of de Gennes narrowing on the wave-vector dependence~\cite{vanMegen}.  A derivation of the strain term is provided in the Supplmentary Material.  The solid lines in Fig.~\ref{figCorrFit}(b) are the results of fits using~\eqref{eq_fitmodel} with  $\Delta r$ and $\Delta r_\text{diff}^2$ as free parameters, illustrating that $\tilde{C}$ at all cycles is well captured by this model. Figure~\ref{figCorrFit}(c) shows $\Delta r$ and $\Delta r_\text{diff}^2$ as functions of cycle number at $\gamma_\text{train}$ = 8\%. The magnitudes of both kinds of irreversible displacements are large initially and decrease with increasing cycle number before reaching small, constant values. This evolution demonstrates the effect of the cyclic shearing, where changes in the microstructure lead the particles to new configurations that are less susceptible to further shear-induced changes, thereby reducing the displacements in subsequent cycles.

The identification of heterogeneous strain fields as a significant component of the irreversible displacements under cyclic shear is a key finding of this work.  While previous research has focused on irreversible particle rearrangements associated with local yielding events that are typically characterized by diffusion-like mean squared displacements, the results here show that heterogeneous strain represents a second major aspect of the shear-induced dynamics that becomes trained to undergo increased reversibility.  In fact, these irreversible strains are likely closely linked to the local yielding, comprising perhaps the Eshelby-like strain fields generated around local yielding events~\cite{JensenPRE2014,NicolasRMP2018}.  As the local yielding events become sparser with increasing cycle number, the associated average strain displacement decreases.  The high sensitivity of XPCS for observing strain in amorphous materials~\cite{LehenyCurOpn2015,ChenPRM2020} has allowed the measurements to bring this contribution to the fore.

The small steady-state displacement amplitudes reached at large cycle number in Fig.~\ref{figCorrFit}(c) signify that the microstructure becomes well trained and reaches a limit to support reversible trajectories for particles under a particular macroscopic strain. 
The essentially reversible microstructure after a large number of cycles is also consistent with previous simulations on a binary glass and experiments on two-dimensional glasses, which show the training leads to the particles taking closed trajectories during one macroscopic strain cycle~\cite{FioccoPRL2014}. Note $\Delta r_\text{diff}^2$ in Fig.~\ref{figCorrFit}(c) falls below zero at $N =$ 5 and 6. We attribute this result to an overestimation of the contribution of thermal dynamics in the normalization by Eq.~\eqref{eq_normalizedC}. Since thermally induced cage dynamics and shear-induced diffusion affect $C(\textbf{q},t_1,t_2)$ with the same wave-vector dependence, an overestimation of the localization length associated with caged motion would lead to an underestimation of $\Delta r_\text{diff}^2$. The estimate of the thermal contribution is based on measurements of the quiescent glass prior to the cyclic shearing, suggesting changes in the microstructure during the training make the fast, thermally driven caged motion of the particles more constrained than in the quiescent glass.

\begin{figure}
    \centering
    \includegraphics[width = 2.5in]{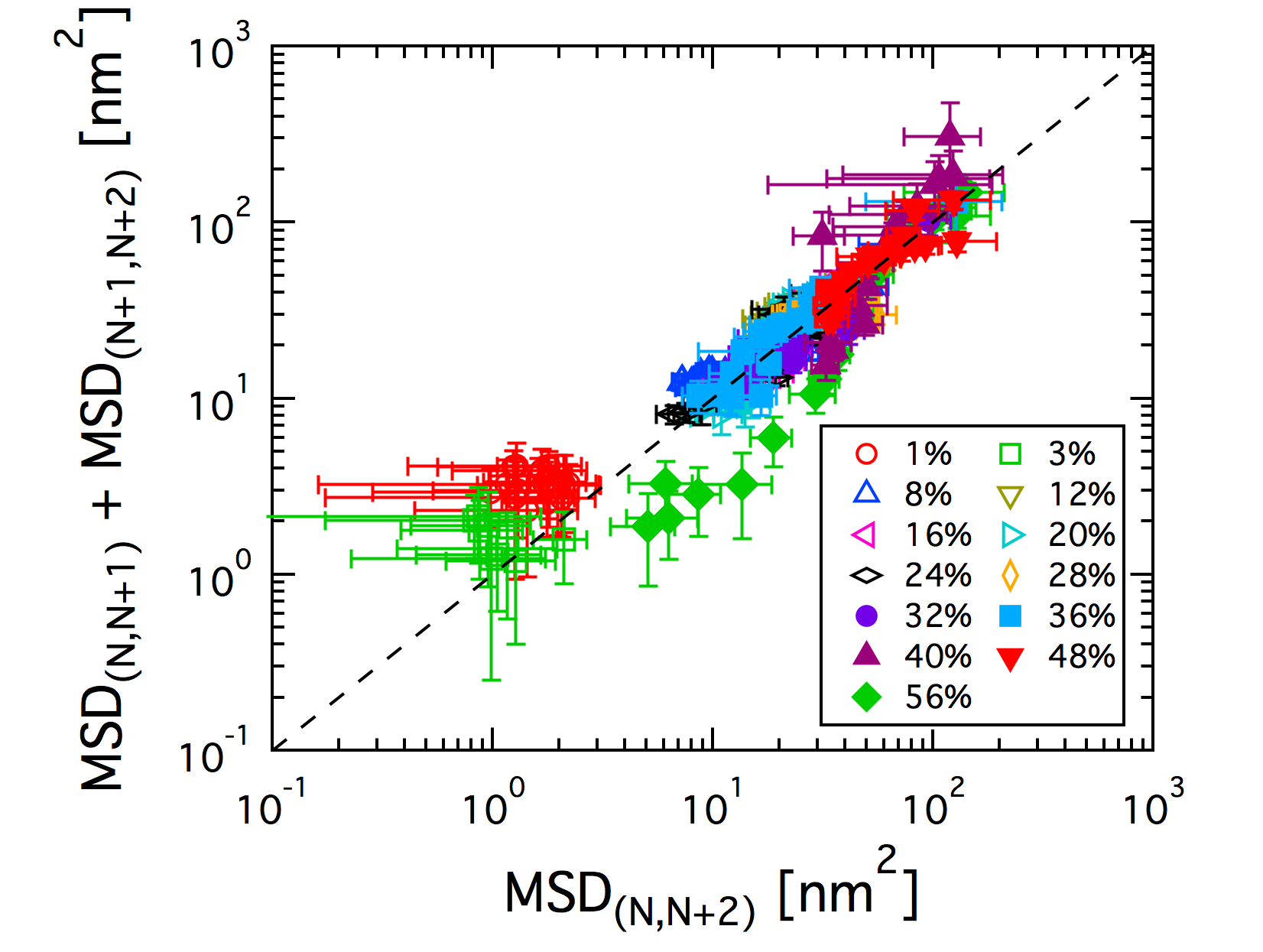}
    \caption{\textbf{Random-walk-like nature of shear-induced displacements.}  The sum of the MSDs in two consecutive cycles is compared with the total MSD during the two cycles from measurements with $\gamma_\text{train}$ specified in the legend. The dashed line is identity.}
    \label{figDisRelation}
\end{figure}

 Combining the diffusive and strain contributions, one can identify a total mean squared irreversible displacement,
\begin{equation}
    \text{MSD} = \Delta r_\text{diff}^2 + (\Delta r)^2,
    \label{eq_defMSD}
\end{equation}
 We denote $\text{MSD}_{(i,j)}$ as the mean squared displacement between microstructures following the $i$th and $j$th cycles. Figure~\ref{figDisRelation} shows the relation between $\text{MSD}_{(N,N+1)}$ + $\text{MSD}_{(N+1,N+2)}$ and $\text{MSD}_{(N,N+2)}$, where the displacements over two cycles are extracted through Eq.~\eqref{eq_fitmodel} using $\tilde{C}(q;N,N+2)$, the normalized correlation function of regions centered two periods from the diagonal ($|t_1 - t_2| = 2T$). The figure includes data from measurements with $\gamma_\text{train}$  from 1\% to 56\% and from all cycles $N$ measured. The displacements behave like a random walk,
\begin{equation}
    \text{MSD}_{(N,N+2)} = \text{MSD}_{(N,N+1)} + \text{MSD}_{(N+1,N+2)},
\end{equation}
indicating the directions of the displacements in consecutive cycles are unbiased random and independent of each other.

\begin{figure*}
    \centering
    \includegraphics[width = 6.5in]{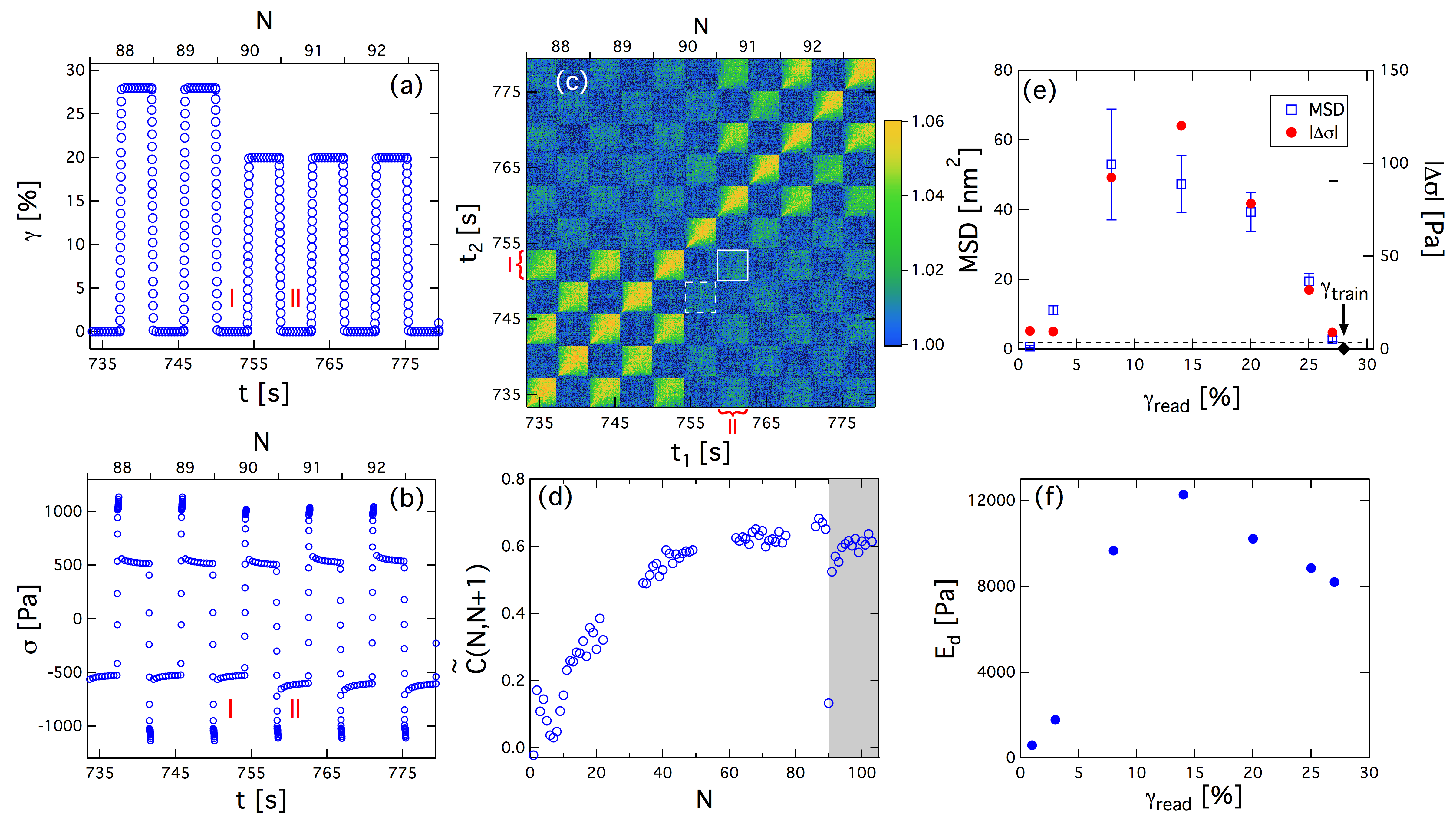}
    \caption{\textbf{Memory of the training strain revealed through microstructural and rheological measurements.} (a) Applied strain as a function of time in a memory experiment during the switch from $\gamma_\text{train}$ = 28\% to $\gamma_\text{read}$ = 20\% following 89 cycles at $\gamma_\text{train}$.  (b) Corresponding stress as a function of time.  (c) $C(\textbf{q},t_1,t_2)$ at $q$ = 0.37 nm$^{-1}$ in the vorticity direction during the switch from $\gamma_\text{train}$ to $\gamma_\text{read}$.  The Roman numbers in (a), (b), and (c) indicate the same intervals of fixed strain.  The dashed white box in (c) demarcates the correlations between the final inverval at $\gamma_\text{train}$ = 28\% and the first interval at $\gamma_\text{read}$ = 20\%.  The solid white box demarcates the correlations between the intervals at $\gamma =$ 0\% immediately succeeding the final inverval at $\gamma_\text{read}$ = 28\% and first interval at $\gamma_\text{read}$ = 20\%.
    (d) $\tilde{C}(q;N,N+1)$ as a function of cycle number during the measurement. The shadowed region indicates the reading cycles. (e) Mean squared displacement (left axis) and the absolute change in average stress (right axis) during the first reading cycle following training at $\gamma_\text{train}$ = 28\% as a function of reading strain amplitude. The arrow indicates $\gamma_\text{train}$ = 28\%. The dashed line indicates the MSD of a training cycle in a well trained system with $\gamma_\text{train}$ = 28\%. The diamond at 28\% represents the change in stress after the last training cycle.  (f) The dissipation during the first reading cycle as a fucntion of reading strain amplitude.}  
    \label{figMemoryMeasurement}
\end{figure*}

\section{Memory}

The training process described above alters the microstructure of the soft glass such that it is more reversible in reponse to the cyclic shear.  Memory of the strain history encoded by this training can be read by applying cyclic shear of varying amplitude $\gamma_\text{read}$~\cite{FioccoPRL2014,KeimRMP2019}. 
Figure~\ref{figMemoryMeasurement}(a), which shows the strain as a function of time during such a memory readout experiment with $\gamma_\text{train}$ = 28\% and $\gamma_\text{read}$ = 20\%, illustrates the process. Initially, 89 cycles of strain at $\gamma_\text{train}$ = 28\% were imposed, then the strain amplitude was reduced to $\gamma_\text{read}$ = 20\%.  Figure~\ref{figMemoryMeasurement}(b) shows the corresponding stress response, and Fig.~\ref{figMemoryMeasurement}(c) shows the colormap of $C(\textbf{q},t_1,t_2)$ at $q$ = 0.37 nm$^{-1}$ during the final cycles at 28\% and the first cycles at 20\%. The colormap has the familiar features of high correlation regions centered along the diagonal and low correlation regions centered at odd numbers of half periods from the diagonal ($|t_1 - t_2| = T/2$, $3T/2$, $5T/2$, ...) like in Figs.~\ref{figTrainingCorr}(a) and (b). However, the square regions of $C(\textbf{q},t_1,t_2)$ centered at one period from the diagonal ($|t_1 - t_2| = T$) show a unique evolution upon the beginning of the reading cycles.  Specifically, before the first reading cycle near $t$ = 754 s, the regions show high correlation, indicating the system has been well trained at 28\% strain to achieve high microstructural reversibility between adjacent cycles. The microstructure during the first interval at $\gamma_\text{read}$ = 20\% shows low similarity with that from the previous cycle due to the difference in macroscopic strain, as shown by the low correlation region bounded by 754 s $ < t_1 < $ 758 s and 746 s $< t_2 <$ 750 s and demarcated by the dashed white box in Fig.~\ref{figMemoryMeasurement}(c). In addition, the two intervals at zero strain before and after the first reading period, which are labeled as I and II, respectively, in Fig.~\ref{figMemoryMeasurement}(a) and (c), also have low correlation, as indicated by the region in the solid white box in Fig.~\ref{figMemoryMeasurement}(c). This low correlation implies large irreversible displacements during the first cycle at 20\% strain despite the fact that the sample was well trained at 28\% strain. As the number of reading cycles increases, the correlations centered at one period from the diagonal ($|t_1 - t_2| = T$) increase, indicating the glass microstructure is now being trained at $\gamma_\text{read}$.


This behavior is further illustrated by $\tilde{C}$, as shown in Figure~\ref{figMemoryMeasurement}(d).  During the initial training stage at $\gamma_\text{train}$ = 28\%, $\tilde{C}$ increases from zero and reaches a plateau in a manner similar to the trend at 36\% strain in Fig.~\ref{figTrainingCorr}(c). After the first cycle of the reading stage at $\gamma_\text{read}$ = 20\%, $\tilde{C}$ decreases significantly and then gradually increases again before reaching a plateau due to training at the new strain amplitude. The dramatic drop in $\tilde{C}$ after the change in strain amplitude shows that a well trained microstructure at one strain amplitude is not trained for high reversibility at a smaller strain, which is consistent with previously results from simulation~\cite{FioccoPRL2014}. In the mean time, this large loss of correlation enables a method to read out the encoded training amplitude by comparing the magnitude of rearrangements after a reading cycle~\cite{KeimRMP2019,FioccoPRL2014}.


 Figure~\ref{figMemoryMeasurement}(e) shows the MSD during the first reading cycle as a function of $\gamma_\text{read}$ for $\gamma_\text{train}$ = 28\%. The MSD has a non-monotonic relation with the reading amplitude, having two minima near zero and $\gamma_\text{train}$, consistent with previous simulations and experiments on 2D colloidal systems~\cite{FioccoPRL2014,KeimPRR2020}. The dashed line in Fig.~\ref{figMemoryMeasurement}(e) is the MSD of a training cycle on a well trained sample, which represents a baseline for encoding the training strain. This non-monotonic behavior indicates that the particles in a well trained sample during a training cycle do not follow the same paths during the ascending and descending half-cycles of strain but instead trace out loops.  As a result, when a smaller strain amplitude is applied, the particles stop at an intermediate location on the loop during the ascending half and hence do not stay on their trained trajectories during the descending half, leading to irreversible displacements~\cite{LundbergPRE2008,KeimPRL2014,KeimSciAdv2021}.  Further, since the scattering wave vector $\bf{q}$ is in the vorticity direction, the MSDs in Fig.~\ref{figMemoryMeasurement}(e) are measures of the irreversible displacements in that direction, demonstrating that the loop trajectories are not confined to the flow-gradient plane but extend into the vorticity direction of the shear. 


The memory of the training amplitude is also seen in the evolution of the glass's macroscopic residual stress and dissipation.  First, during the memory readout stage the stress does not continue the steady state behavior achieved during the training stage after the first reading cycle but instead becomes quantitatively different by an amount that depends on $\gamma_\text{read}$.  This difference is illustrated by the different plateaus in stress reached in the intervals labelled I and II in Fig.~\ref{figMemoryMeasurement}(b).  Figure~\ref{figMemoryMeasurement}(e) shows the absolute change in the average stress $|\Delta\bar{\sigma}|$ during the interval of zero strain immediately before and after the first reading cycle as a function of $\gamma_\text{read}$. The diamond represents the change in stress after a cycle in the steady training stage, which is very close to zero. Similar to the MSD, $|\Delta\bar{\sigma}|$ has a non-monotonic relation with $\gamma_\text{read}$, with $|\Delta\bar{\sigma}|$  having two minima near zero and $\gamma_\text{train}$ and a maximum near $\gamma_\text{read} = \gamma_\text{train}/2$. The behavior of the macroscopic stress hence mirrors that of the microscopic MSD in both the training and memory readout measurements, demonstrating that the macroscopic residual stress can serve as another feature to decode the memory of shear history in a glass.  Similarly, the dissipation in the first cycle at $\gamma_\text{read}$ shows a non-monotonic variation with $\gamma_\text{read}$ with a peak at $\gamma_\text{read} = \gamma_\text{train}/2$, as shown in Fig.~\ref{figMemoryMeasurement}(f).  Notably, this signature of the memory of $\gamma_\text{train}$ in the energy dissipated is remarkably similar to that reported from experiments on steel wool, treated as a model three-dimensional frictional amorphous material, subjected to cyclic compressional strain~\cite{ShohatPRL2023}.

\section{Conclusion}

This work has revealed several new key aspects of memory formation and readout in glasses through cyclic shear.  First, by characterizing the phenomena in a bulk, thermal glass, particularly its 3D nature, the results demonstrate behavior seen previously in experiment only in 2D glasses and connect it to phenomena in other disordered systems that can be trained and display memory, such as gels~\cite{SchwenSofMat2020} and granular material~\cite{BensonPRE2021}.  Second, as illustrated in Fig.~\ref{figMemoryMeasurement}, the measurements extend the observations of memory in the glass beyond microstructural features to the rheological behavior, specifically the residual stress and dissipation, thus illustrating how memory can become encoded in a macroscopic material property. Third, as mentioned above, while previous research has focused on irreversible particle rearrangements associated with local yielding events, the results here show that non-affine strain fields represent a second major aspect of the shear-induced dynamics that becomes trained to undergo increased reversibility.  Importantly, these heterogeneous strains are directly related to the heterogeneous internal stress in the glass and hence to the residual stress measured in rheology.  While the importance of training and memory of the local particle rearrangments to the material properties of glasses is unclear, the demonstration of training and memory in the heterogeneous strains provides a clear connection between the microscopic and macroscopic manifestations of training and in glasses.  That is, our findings show how microscopic memory impacts the mechanical behavior of glasses.  Future work that builds on this connection would be valuable for understanding and ultimately controlling rheological memory in out-of-equilibrium materials.

\section{Materials and Methods}

\subsection{Nanocolloidal soft glass}
The colloidal glass was composed of a bidisperse suspension of spherical charged silica particles with diameters of 26 nm (Ludox TM50, Sigma Aldrich) and 12 nm (Ludox HS40, Sigma Aldrich) in water. The original volume fraction of the aqueous suspensions of the large and small particles are 30\% and 22\%, respectively, according to the manufacturer, and the particles are stabilized by negatively charged surfaces and counterions in the suspension. The two suspensions were first mixed at a silica volume ratio of 1:4 (big to small particles). Then, 11 mL of the mixture was loaded into a dialysis cassette (3.5K MWCO, Thermo Fisher), and the cassette was immersed in a solution of dextran (MW ca 40,000, Alfa Aesar) with a concentration of 0.6 g/mL for 2.5 hours to concentrate the mixture through dialysis. The final silica volume fraction of the suspension was $\phi$ = 39\%, obtained by comparing the weight of a portion of the sample before and after drying under vacuum at 40 C. The resulting sample was a ductile soft glass. By ``soft'' we refer to the repulsive particles interactions, which were a screened Coulomb interaction.

\subsection{Rheo-XPCS}
X-ray photon correlation spectroscopy and \textit{in situ} rheology measurements were carried out at Sector 8ID-I of Advanced Photon Source. 0.3 mL of sample was loaded in a Couette cell of a stress-control rheometer (Anton Paar MCR301) mounted along the beam path. The cell was made of thin-walled polycarbonate with inner (bob) diameter of 11 mm and outer (cup) diameter of 11.4 mm. A partially coherent x-ray beam with energy 10.9 keV was focused from a size of 150$\times$15 $\mu$m$^2$ (V$\times$H) to 3$\times$15 $\mu$m$^2$ (V$\times$H) at the sample. The incident beam transited horizontally through the center of the cell whose axis was oriented vertically. An area detector (X-spectrum LAMBDA 750 K)~\cite{PennicardJourInstr2011,KhanJSR2018} 4 meters downstream of the cell detected the scattered x-rays over a wave-vector range of 0.03 nm$^{-1} < q <$ 0.68 nm$^{-1}$ at a rate of 50 frames per second. X-ray images were obtained continuously during the measurement except for brief periods every approximately 200 seconds while the sample was translated slightly to limit the radiation exposure to any one region (leading, e.g., to the gap in the data near $N = 30$ in Fig.~\ref{figTrainingCorr}(c)). Since the incident beam was parallel to the local flow gradient direction in the Couette cell, in the small angle scattering regime, the scattered wave-vector $\textbf{q}$ was in the flow-vorticity plane.

Prior to each measurement, the sample was sheared at a rate of $\dot{\gamma}$ = 1 $s^{-1}$ for 5 min and then held at zero stress for 5 min to reset the sample state. In the training measurements, cyclic shear strain with a rectangular wave form was applied to the sample with peak to peak strain amplitudes $\gamma_\text{train}$ ranging from 1\% to 56\%. During each cycle, the strain was first held at zero for 4 s before ramping to $\gamma_\text{train}$ in 0.2 s, then the strain was held at $\gamma_\text{train}$ for 4 s before ramping back to zero in 0.2 s, resulting in a period of $T$ = 8.4 s. To investigate the memory of the shear history, the sample was first trained with $\gamma_\text{train}$ = 28\% for 90 cycles, and then rectangular-wave strain cycles with a reading amplitude $\gamma_\text{read}$ ranging from 1\% to 27\% were applied.

 \section{Acknowledgements}
We thank D. Shohat and Y. Lahini for helpful correspondence.  Funding was provided by the NSF (CBET-1804721) and the NSERC Discovery grant program.  The research used resources of the Advanced Photon Source and the Center for Nanoscale Materials, U.S. Department of Energy (DOE) Office of Science User Facilities operated for the DOE Office of Science by Argonne National Laboratory under Contract No. DE-AC02-06CH11357

\end{document}


\title{Supplementary Material for \\``Microstructural and rheological training and memory of nanocolloidal soft glasses under cyclic shear''}

\author{Yihao Chen}
\affiliation{%
Department of Physics and Astronomy, Johns Hopkins University, Baltimore, Maryland 21218, USA
}%
\author{Simon A. Rogers}%
\affiliation{%
Department of Chemical and Biomolecular Engineering, University of Illinois Urbana-Champaign, Champaign, IL 61801, USA
}%
\author{Suresh Narayanan}
\affiliation{X-Ray Science Division, Argonne National Laboratory, Argonne, Illinois 60439, USA}
\author{James L. Harden}
\affiliation{%
 Department of Physics, University of Ottawa, Ottawa, Ontario K1N 6N5, Canada
}%
\author{Robert L. Leheny}
\affiliation{%
Department of Physics and Astronomy, Johns Hopkins University, Baltimore, Maryland 21218, USA
}%

\maketitle

\section{Stress response to step strain}
\begin{figure}
    \centering
    \includegraphics[width = 3.1in]{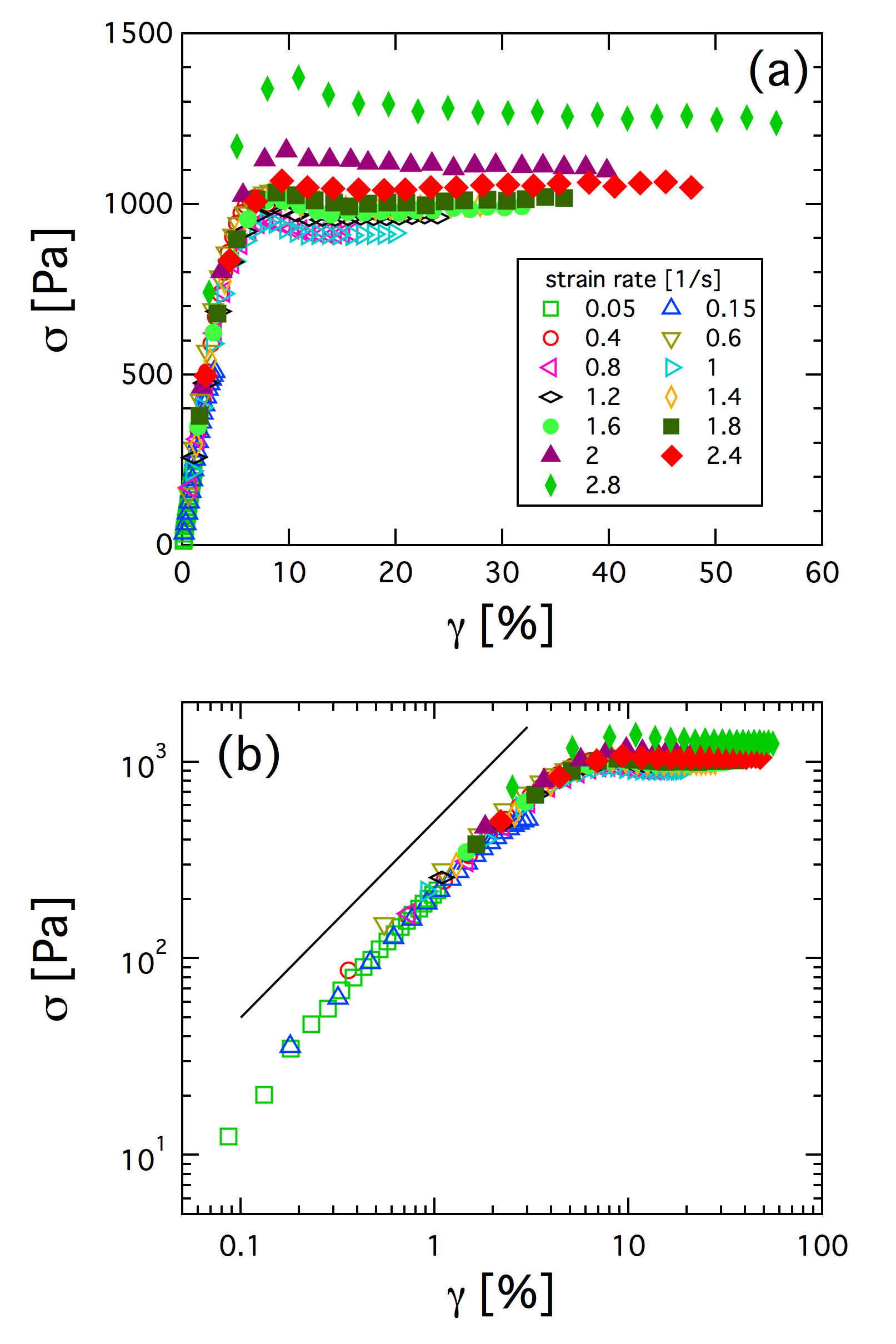}
    \caption{Stress as a function of strain on a (a) linear and (b) log scale during the initial ramp in strain at the commencement of training measurements at various strain amplitudes from 1\% to 56\%.  In all cases, the ramp time was 0.2 s, leading to the different strain rates indicated in the figure legend.  The solid line depicts a linear relation between stress and strain.}
    \label{stepstrain}
\end{figure}

Figures~\ref{stepstrain} (a) and (b) show the stress as a function of strain on linear and log scales, respectively, during the initial ramp in strain at the commencement of training measurements at various strain amplitudes from 1\% to 56\%.  In all cases, the ramp time was 0.2 s, leading to the different strain rates indicated in the figure legend.  The stress is linear in the strain below approximately 2\% strain.  Above aproximately 10\% strain, the stress reaches a plateau that depends on strain rate.

\section{Instantaneous Correlation Function in the flow direction during training}

\begin{figure}
    \centering
    \includegraphics[width = 3.1in]{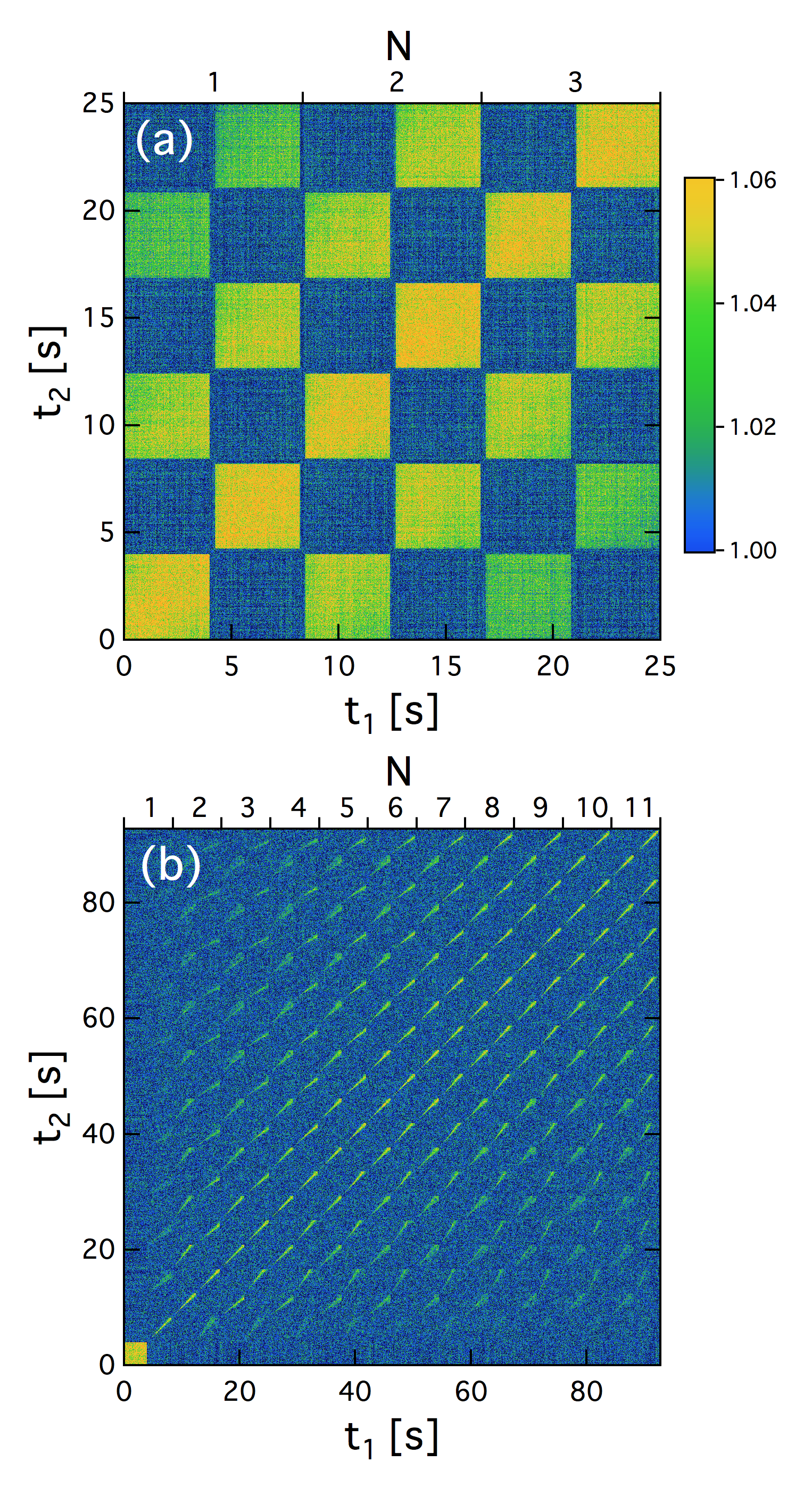}
\caption{(a) $C(q, t_1, t_2)$ during the first three cycles of a measurement with $\gamma_\text{train} =$ 1\%. (b) $C(q, t_1, t_2)$ during the first 11 cycles of a measurement with $\gamma_\text{train} =$ 8\%. (a) and (b) share the colorbar. In both cases, $q = 0.37$ nm$^{-1}$ in the flow direction.}
    \label{twotimeflow}
\end{figure}

Figures~\ref{twotimeflow} (a) and (b) show colormaps of $C(q, t_1, t_2)$ during the first few cycles of measurements with $\gamma_{\text{train}} = 1\%$ and $8\%$, respectively, at $q = 0.37$ nm$^{-1}$ parallel to the flow directon.  The results are from the same measurements as in Figs.~1(a) and (b) in the manuscript.  The effects of stress relaxation during periods of fixed strain~\cite{ChenPRM2020} are pronounced at $8\%$ strain.

\section{Structure of binary nanocolloidal glass}

\begin{figure}
    \centering
    \includegraphics[width = 3.1in]{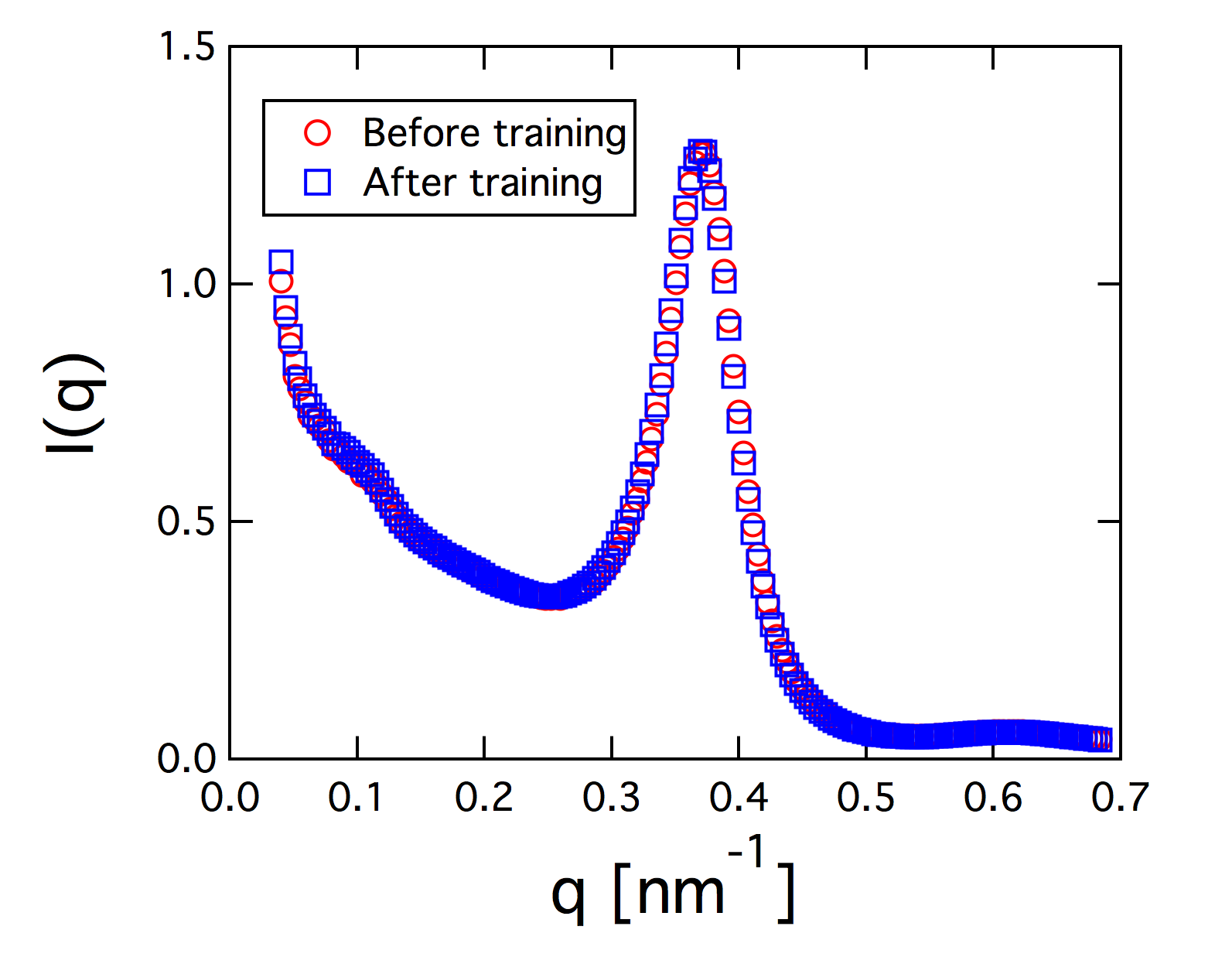}
    \caption{Scattering intensity as a function of the magnitude of the scattering wave vector before (circles) and after (squares) all of the training measurements conducted in the experiment.}
    \label{figStrainsweepIq}
\end{figure}

Figure~\ref{figStrainsweepIq} shows the x-ray scattering intensity $I(q)$ measured before and after all of the training measurements conducted in the experiment. $I(q)$ has a large peak near $q$ = 0.37 nm$^{-1}$ corresponding to a structure-factor peak associated with near neighbor spatial correlations of the smaller particles in the glass. The identical shape of $I(q)$ before and after the training measurements indicates that no observable demixing or other changes to the sample occurred during the experiment.

\section{Siegert factor}
\begin{figure}
    \centering
    \includegraphics[width = 3.1in]{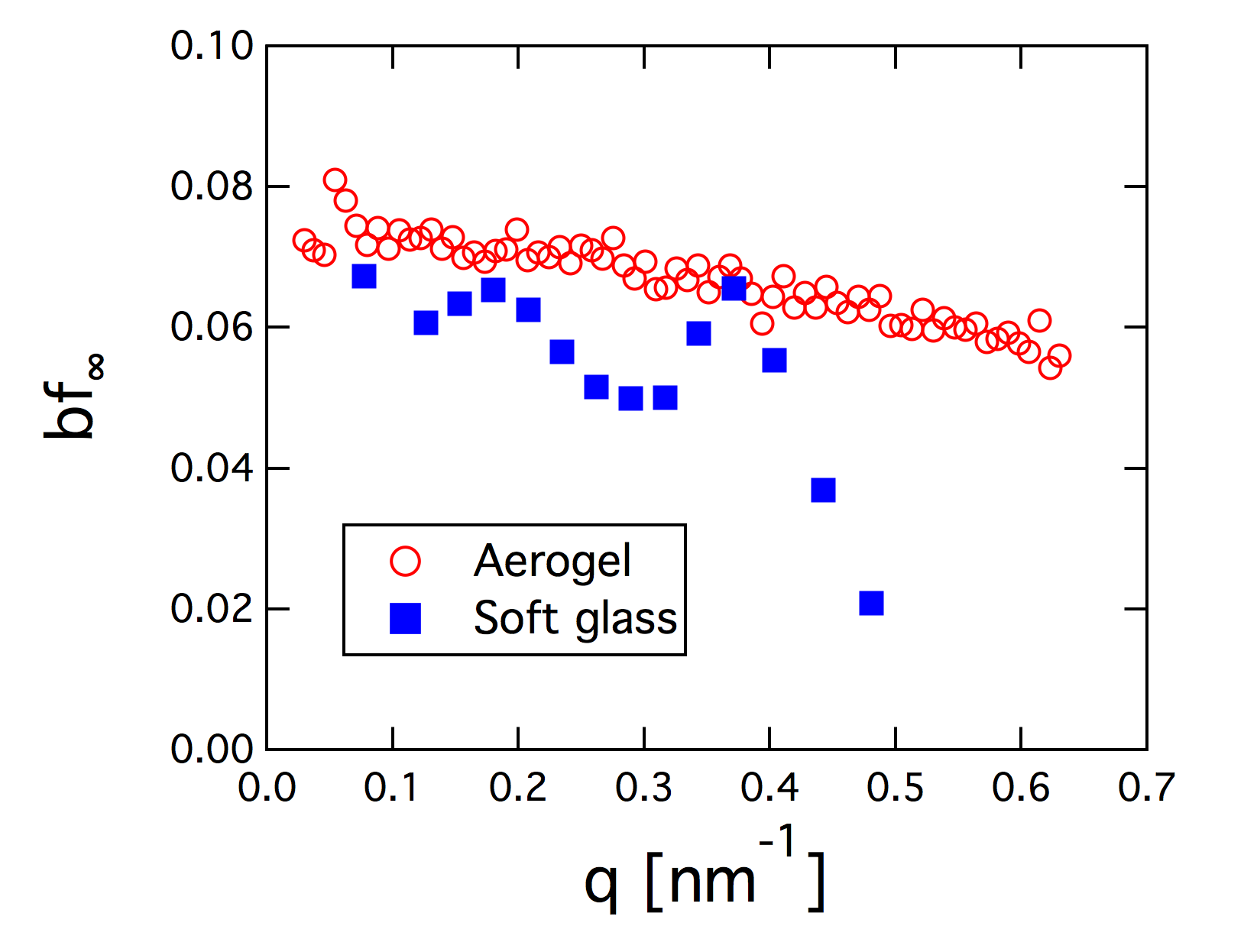}
    \caption{The product $bf_{\infty}$ (squares) obtained from the amplitude of the XPCS correlation function, $g_2(\mathbf{q},\tau) -1$, at small $\tau$ measured on the soft glass.  Also shown is an estimate of the Siegert factor $b$ (circles) determined from a separate XPCS measurement on a thin aerogel sample, as explained in the text.  The wave-vector-dependent non-ergodicity parameter $f_{\infty}$ is a measure of the fourier components of the arrested concentration fluctuations in the glass~\cite{vanMegen}.}
    \label{Siegertfactor}
\end{figure}

 For pairs of speckle patterns taken at fixed strain and small values of $|t_1-t_2|$,  $C(q_v,t_1,t_2) \approx bf_{\infty}+1$, where $b$ is the Siegert factor that depends on instrumental factors such as the coherence of the incident beam, the solid angle subtended by a detector pixel, and the sample thickness, and $f_{\infty}$ is the wave-vector-dependent non-ergodicity parameter of the glass~\cite{vanMegen}. Results for $bf_{\infty}$ as a function of $q$ are shown in Fig.~\ref{Siegertfactor}.  To obtain an estimate of the Siegert factor, we measured $C(q_v,t_1,t_2)$ at small $|t_1-t_2|$ for a thin Aerogel sample.  Because in the rheo-XPCS measurements the scattering from the front and back of the Couette cell add primarily incoherently, the effective Siegert factor is reduced approximately by a factor of two. (See Methods section the manuscript.) The data shown for Aerogel in Fig.~\ref{Siegertfactor} are the weakly $q$-dependent values of $b$ measured with the Aerogel divided by two, and hence can be taken as an approximation, but slight underestimation, of $b$ in the experiment.

\section{Derivation of Strain Term}

In this section, we provide a derivation of the contribution to the normalized correlation function in Eq.~(4) of the manuscript due to irreversible strain. 
 Previous simulation and 2D experiments on colloidal glass under cyclic shear have shown that the plastic deformation occurs in localized regions that become increasingly sparse as the glass becomes trained.  These regions have the character of an Eshelby inclusion that drives quadrupolar elastic deformation of the surrounding material~\cite{Eshelby1957,NicolasRMP2018,KeimSciAdv2021}, leading to particle displacements $u$ outside a plastic region that (in 3D) vary with distance $r$ from the region to leading order as~\cite{Picard,Schall,JensenPRE2014}
\begin{equation}
    u(r) \sim r^{-\alpha}
\end{equation}
with $\alpha = 3$.  (The deformation field about a single region is anisotropic, but the scattering volume in the XPCS experiment presumably contains many regions, so we assume an average over orientations.) Given that these displacements occur over a cycle of period $T$, we can say that the particles have an effective time-average strain velocity $V = u/T$, so that
\begin{equation}
    V(r) \sim r^{-\alpha}
     \label{Vofr}
\end{equation}
The number of particles $dN$ at a distance from a plastic region between $r$ and $r+dr$ is $dN \sim r^2dr$.  Combining this with Eq.~(\ref{Vofr}) leads to 
\begin{equation}
    \frac{dN}{dV} \sim V^{-(3/\alpha +1)}.
\end{equation}
$dN/dV$ in turn is proportional to the velocity probability density $W(V)$, where $W(V)dV$ is the probability that the magnitude of a particle's strain velocity is between $V$ and $V+dV$.  The dynamic structure factor from such a distribution of velocities is~\cite{Berne,CipellettiFaraday},
\begin{equation}
    f(q,t) = \int_0^{\infty} dV W(V) \frac{\sin(qVt)}{qVt}
    \label{fqt}
\end{equation}
For a velocity probability density with a power-law tail like Eq.~(\ref{Vofr}), Cipelletti and coworkers have further shown that Eq.~(\ref{fqt}) leads to~\cite{CipellettiFaraday}
\begin{equation}
    f(q,t) = \exp\left[-(V_0qt)^{3/\alpha}\right],
    \label{compressedexpon}
\end{equation}
where $V_0$ is a time-averaged characteristic velocity.  For the strain in the material around the plastic deformations for which $\alpha = 3$, this result corresponds to
\begin{equation}
    f(q,t) = \exp\left[-V_0qt\right].
    \label{sixfifths}
\end{equation}
The intensity autocorrelation function measured in the XPCS measurement is proportional to one plus the square of $f(q,t)$, which leads to a strain contribution to the normalized correlation function from a cycle of shear of
\begin{equation}
\tilde{C}_{strain}(q) = \exp\left[-q\Delta r\right],
\label{Ctilda}
\end{equation}
where $\Delta r = 2V_0 T$ is a characteristic strain displacement.

\bibliography{training_references}